# Comments on Five Smart Card Based Password Authentication Protocols

Yalin Chen
Institute of Information Systems and Applications, NTHU, Tawain
d949702@oz.nthu.edu.tw

Jue-Sam Chou*
Dept. of Information Management
Nanhua University, Taiwan
jschou@mail.nhu.edu.tw
*: corresponding author

Chun-Hui Huang
Dept. of Information Management
Nanhua University, Taiwan
g6451519@mail.nhu.edu.tw

*Abstract¡* In this paper, we use the ten security requirements proposed by Liao et al. for a smart card based authentication protocol to examine five recent work in this area. After analyses, we found that the protocols of Juang et al.¡s , Hsiang et al.¡s, Kim et al.¡s, and Li et al.¡s all suffer from offline password guessing attack if the smart card is lost, and the protocol of Xu et al.¡s is subjected to an insider impersonation attack.

## I. INTRODUCTION

Password authentication protocols have been widely adopted for a user to access a remote server over an insecure network. In recent, many smart card password authentication protocols [1-20] are proposed, which emphasizes two-factor authentication mechanism to enhance the user end¡s security. One factor is the user-rememberable password while the other factor is the user-possessing smart card which is a tamper-resistant device with storage and computational power. Moreover, recent studies investigated a weakness of a traditional password authentication protocol. That is, in the traditional one the server usually maintains a password or verification table to store user authentication data. However, this approach will make the system easily subjected to impersonation or stolen-verifier attack if the table is compromised.

In 2006, Liao et al. [2] identified ten security requirements to evaluate a smart card based password authentication protocol. We show them as follows.

R1. It needs no password or verification table in the server.

R2. The client can choose and change his password freely.

R3. The client needs not to reveal their password to the server even in the registration phase.

R4. The password should not be transmitted in plaintext over the network.

R5. It can resist insider (a legal user) attack.

R6. It can resist replay attack, password guessing attack, modification-verification-table attack, and stolen-verifier attack.

R7. The length of a password should be appropriate for memorization.

R8. It should be efficient and practical.

R9. It should achieve mutual authentication.

R10. It should resist offline password guessing attack even if the smart card is lost.

In their article, they also proposed a protocol to satisfy these ten security requirements. But Xiang et al. [9] demonstrated that their protocol suffers from both the replay attack and the password guessing attack. Other than theirs, many efforts trying to propose a secure protocol were made recently. For example in 2008, Juang et al. [7] proposed an efficient password authenticated key agreement using bilinear pairings. In 2009, Hsiang et al. [14], Kim et al. [16], and Xu et al. [18] each also proposed a protocol of this kind, respectively. In this year 2010, Li et al.[20] also proposed a protocol in this area. Although they claimed their protocols are secure. However, in this paper, we will show some weaknesses in [18], [7], [14], [16], [20], correspondingly.

The remainder of this paper is organized as follows: In Section II, we review and attack on the scheme of Juang et al.¡s [7]. Then we review and attack on the protocols of Hsiang et al. ¡s [14], Kim et al. [16], Xu et al. ¡s [18], and Li et al. ¡s [20] in Section III through VI, respectively. Finally, a conclusion is given in Section VIII.

## II. REVIEW AND ATTACK ON JUANG ET AL.'S SCHEME

In their scheme [7], if an attacker gets C¡s smart card, he can successfully launch an offline password guessing attack. Hence, the scheme cannot satisfy requirement R10. In the following, we first review Juang *et al.*¡s protocol and then show the attack on the protocol.

### A. Review

Their protocol consists of four phases: the setup phase, the registration phase, the login and authentication phase, and the password changing phase.

In the setup phase, server S chooses two secrets $s$, $x$ and publishes $P_s = sP$, where $P$ is a generator of an additive cyclic group $G_1$ with a prime order $q$. S also publish a secure hash function H(¡).

In the registration phase, user i register his $ID_i$ and H($PW_i$, $b$) to server S. S issues a smart card which contains $b_i$ ($b_i$ = $E_x$[H($PW_i$, $b$), $ID_i$, H(H($PW_i$, $b$), $ID_i$)], $E_x$[M] which is a ciphertext of M encrypted by S¡s secret key $x$), and $b$ (a random number chosen by i).

When i wants to login into S, i starts the login and authentication phase, and sends {$aP$, $\alpha$} to S, where $a$ is a random number chosen by i, $\alpha = E_{Ka}[b_i]$, $Ka$ = H($aP$, $P_s$, $Q$, e($P_s$, $aQ$)), e: $G_1$¡ $G_1 \rightarrow G_2$ is a bilinear mapping, $Q$ = h($ID_s$), h(¡) is a map-to-point hash function, h:{0,1}*$\rightarrow G_1$, and $ID_s$ is S¡s identification. Subsequently, S chooses a random number $r$, computes the session key $sk$ = H(H($aP$, $P_s$, $Q$, e($aP$, $sQ$)), $r$, $ID_i$, $ID_s$) = H($Ka$, $r$, $ID_i$, $ID_s$) since e($P_s$, $aQ$) = e($aP$, $sQ$) , and sends {$Auth_s$, $r$} to user i, where $Auth_s$ = H($Ka$, H($PW_i$, $b$), $r$, $sk$), and H($PW_i$, $b$) is obtained from decrypting $\alpha$ and $b_i$. Then, i computes the session key $sk$. To authenticate S, user i verifies $Auth_s$ to see if it is equal to H($Ka$, H($PW_i$, $b$), $r$, $sk$). If it is, i computes and sends {$Auth_i$} to S, where $Auth_i$ = H($Ka$, H($PW_i$, $b$), $r+1$, $sk$) and H($PW_i$, $b$) is the hash result of $b$ stored in the smart card with $PW_i$ inputted by i. Finally, to authenticating i, S checks to see if $Auth_i$ is equal to H($Ka$, H($PW_i$, $b$), $r+1$, $sk$).

*B. Attack*

In the protocol, supposed that user C lost his smart card and the card is got by an insider E, E can impersonate C to login into S without any detection. We show the attack in the following.

E first reads out $b$ and $b_c$ (which equals $E_x$[H($PW_c$, $b$), $ID_c$, H(H($PW_c$, $b$), $ID_c$)]) stored in C¡s smart card but he doesn¡t have the knowledge of $PW_c$.

In the login and authentication phase, E chooses a random number $c$, computes $cP$, $Kc$ = H($cP$, $P_s$, $Q$, e($P_s$, $cQ$)), $\alpha$ = $E_{Kc}[b_c]$, and sends {$cP$, $\alpha$} to S. After receiving the message, S chooses a random number $r$, computes session key $sk$ = H($Kc$, $r$, $ID_c$, $ID_s$), $Auth_s$ = H($Kc$, H($PW_c$, $b$), $r$, $sk$), and sends {$Auth_s$, $r$} to C. E intercepts the message and launches an off-line password guessing attack as follows.

E chooses a candidate password $PW'$ from a dictionary, computes $Kc$ = H($cP$, $P_s$, $Q$, e($P_s$, $cQ$)), $sk$ = H($Kc$, $r$, $ID_c$, $ID_s$), H($Kc$, H($PW'$, $b$), $r$, $sk$) and checks to see if it is equal to the received $Auth_s$. If it is, the attacker successfully gets C¡s password $PW_c$ which is equal to $PW'$. Subsequently, E can masquerade as C by using $PW'$ and C¡s smart card to log into S. That is, Juang et al.¡s cannot satisfy the security requirement R10: It should resist password guessing attack even if the smart card is lost.

### III. REVIEW AND ATTACK ON THE PROTOCOL OF HSIANG ET AL.'S SCHEME

In this section, we first review Hsiang *et al.*¡s protocol [14] and then demonstrate a smart card lost and offline password guessing attack on the protocol.

*A. Review*

In the protocol, when user C wants to change his password, he inserts his card and types his *ID* and *PW*. The smart card computes $P^* = R \oplus H(b \oplus PW)$, and $V^* = H(P^* \oplus H(PW))$, and compares $V^*$ with $V$, where *PW* is C¡s old password, and $R$, $b$, and $V$ are stored in C¡s smart card. If they are equal, the card verifies user C and accepts his password change request. The card subsequently ask C a new password $PW^*$ and then computes $R_{new} = P^* \oplus H(b \oplus PW^*)$ and $V_{new} = H(P^* \oplus H(PW^*))$. Finally, the card replaces $V$ with $V_{new}$.

*B. Attack*

Assume that an attacker E who gets C¡s smart card, reads the values of $R$, $b$, and $V$, and then launches an offline password guessing attack as follows. E chooses a candidate password $PW'$ from a dictionary, computes $P' = R \oplus H(b \oplus PW')$ and $V' = H(P' \oplus H(PW'))$, and checks to see if $V'$ and $V$ are equal. If they are, $PW'$ is the correct password.

### IV. REVIEW AND ATTACK ON THE PROTOCOL OF KIM ET AL.'S SCHEME

In this section, we first review Kim *et al.*¡s protocol [16] and then demonstrate a smart card lost and offline password guessing attack on the protocol.

*A. Review*

In their protocol, when user C wants to change his password, he inserts his card and types his *ID* and *PW*. The smart card computes $K^*_1 = R \oplus H(PW)$ and compares $K^*_1$ with $K_1$ to see if they are equal, where $R$ (=$K_1 \oplus H(PW_c)$) and $K_1$ (=H($ID \oplus x) \oplus N$ ) are stored in C¡s smart card, $PW_c$ is chosen by the user when he registers himself to the remote server S, and $N$ is a random number. If they are equal, the card verifies user C and accepts his password change request. C subsequently asks C a new password $PW^*$, and then computes $R^* = K^*_1 \oplus H(PW^*)$ and $K^*_2 = K_2 \oplus H(PW \oplus H(PW)) \oplus H(PW^* \oplus H(PW^*))$, where $K_2$ = H($ID \oplus x \oplus N) \oplus H(PW_c \oplus H(PW_c))$ is also stored in C¡s smart card. Finally, the smart card will replace $R$ and $K_2$ with $R^*$ and $K^*_2$, respectively.

*B. Attack*

An attacker E who gets C¡s smart card, reads the values of $R$, $K_1$, and $K_2$, and then launches an offline password guessing attack as follows. E chooses a candidate password $PW'$ from a dictionary, computes $K'_1 = R \oplus H(PW')$, and checks to see if $K'_1$ and $K_1$ are equal. If they are, $PW'$ is the correct password.

### V. REVIEW AND ATTACK ON THE PROTOCOL OF XU ET AL.'S SCHEME

Xu *et al.*¡s protocol [18] can not satisfy security requirements R3 (The client needs not to reveal their password to the server) and R5 (It can resist insider attack). We show the scheme and its violations as follows.

*A. Review*

Xu et al.'s protocol [18] consists of three phases: the registration phase, the login phase, and the authentication phase.

In the registration phase, user C submits his $ID_c$ and $PW_c$ to the server S. S issues C a smart card which stores C's identity $ID_c$, and $B = H(ID_c)^x + H(PW_c)$, where $x$ is S's secret key and $PW_c$ is C's password.

In the login phase, user C inputs $ID_c$ and $PW_c$ to his smart card. The card obtains timestamp $T$, chooses a random number $v$, computes $B_c = (B - H(PW_c))^v = H(ID_c)^{xv}$, $W = H(ID_c)^v$, and $C_1 = H(T, B_c, W, ID_c)$, and sends $\{ID_c, C_1, W, T\}$ to S.

In the authentication phase, after receiving $\{ID_c, C_1, W, T\}$ at time $T^*$, S computes $B_s = W^x$, and checks to see if $ID_c$ is valid, $T^* - T < \Delta T$, and $C_1$ is equal to $H(T, B_s, W, ID_c)$. If they are, S selects a random number $m$, gets timestamp $T_s$, computes $M = H(ID_c)^m$, $C_s = H(M, B_s, T_s, ID_c)$, and sends $\{ID_c, C_s, M, T_s\}$ to C. After receiving the message, C verifies $ID_c$ and $T_s$, computes $H(M, B_c, T_s, ID_c)$, and compares it with the received $C_s$. If they are equal, S is authentic. Then, C and S can compute the common session key as $sk = H(ID_c, M, W, M^v)$ and $sk = H(ID_c, M, W, W^m)$, respectively.

*B. Weaknesses*

First, the scheme obviously violates security requirement R3 since the client transmits clear password in the registration phase.

Second, we show an impersonation attack on the scheme below. Assume that a malicious insider U wants to masquerade as C to access S's resources. He reads $B$ from his smart card, obtains system's timestamp $T_u$, chooses a random number $r$, computes $B_u = (B - H(PW_u))^r = H(ID_u)^{xr}$, $W = H(ID_c)^r$, $C_1 = H(T_u, B_u, W, ID_c)$, and sends $\{ID_c, C_1, W, T_u\}$ to S.

After receiving the message, S validates $ID_c$ and $T_u$, computes $B_s = W^x = H(ID_c)^{rx}$, and checks to see if the received $C_1$ is equal to the computed $H(T_u, B_s, W, ID_c)$. In this case, we can see that $C_1$ is obviously equal to $H(T_u, B_s, W, ID_c)$. Hence, U (who masquerades as C) is authentic. Finally, S obtains timestamp $T_s$ and sends $\{ID_c, C_s, M, T_s\}$ to U, where $M = H(ID_c)^m$ and $m$ is a random number chosen by S. U also can compute the session key as $sk = H(ID_c, M, W, M^r)$ shared with S. Therefore, user U's insider impersonation attack succeeds.

## VI. REVIEW AND ATTACK ON THE PROTOCOL OF LI ET AL.'S SCHEME

In this section, we first review the registration phase, login phase and authentication phase of the protocol in Li et al.'s [20], and then present our attack on the protocol.

*A. Review*

In the registration phase, user C submits his $ID_c$, $PW_c$, and his personal biometric $B_c$ to the server S. S issues a smart card for C, which stores the values of $ID_c$, $f_c = H(B_c)$, and $e_c = H(ID_c, x) \oplus H(PW_c, f_c)$, where $x$ is S's secret key.

In the login phase, user C keys $ID_c$ and $PW_c$ to his smart card and inputs his personal biometric $B_c$ on the specific device to check if $H(B_c)$ is equal to $f_c$ stored in the smart card. If it is, the card selects a random number $R_c$, computes $M_1 = e_c \oplus H(PW_c, f_c) = H(ID_c, x)$, $M_2 = M_1 \oplus R_c$, and sends $\{ID_c, M_2\}$ to S.

In the authentication phase, after receiving $\{ID_c, M_2\}$, S checks to see if $ID_c$ is valid. If it is, S chooses a random number $R_S$, computes $M_3 = H(ID_c, x)$, $M_4 = M_2 \oplus M_3 = R_c$, $M_5 = M_3 \oplus R_S$, $M_6 = H(M_2, M_4)$, and sends $\{M_5, M_6\}$ to C. After receiving S's message, C verifies whether $M_6$ is equal to $H(M_2, R_c)$. If it is, S is authentic. C then computes $M_7 = M_5 \oplus M_1 = M_3 \oplus R_S \oplus M_1 = H(ID_c, x) \oplus R_S \oplus H(ID_c, x) = R_S$, $M_8 = H(M_5, M_7)$, and sends $\{M_8\}$ to S. After receiving C's message, S verifies whether $M_8$ is equal to $H(M_5, R_s)$. If it is, C is authentic. S then accepts C's login request.

*B. Attack*

Assume that an attacker E gets C's smart card and reads the values of $ID_c$, $f_c$ and $e_c$. He can launch an offline password guessing attack by sending only one login request to the server. We show the attack as follows.

E chooses a random number $M_e$ and sends $\{ID_c, M_e\}$ to S. After receiving the message, S checks to see if $ID_c$ is valid. If it is, S chooses a random number $R_S$, computes $M_3 = H(ID_c, x)$, $M_4 = M_e \oplus M_3$, $M_5 = M_3 \oplus R_S$, $M_6 = H(M_e, M_4)$, and sends $\{M_5, M_6\}$ to E. After receiving S's message, E terminates the communication, chooses a candidate password $PW'$ from a dictionary, computes $M' = H(M_e, M_e \oplus e_c \oplus H(PW', f_c))$, and compares to see if $M'$ is equal to $M_6$. If they are, $PW'$ is the correct password, since $M_e \oplus e_c \oplus H(PW', f_c) = M_e \oplus H(ID_c, x) \oplus H(PW_c, f_c) \oplus H(PW', f_c)$. If $PW' = PW_c$, then the equation equals to $M_e \oplus H(ID_c, x)$ which equals to $M_e \oplus M_3 = M_4$. That is, $M' = H(M_e, M_4) = M_6$.

## VII. CONCLUSION

Smart-card based password authentication protocols provide two-factor authentication mechanism to improve the user end's security than the traditional ones. Liao et al. proposed ten security requirements to evaluate this kind of protocols. According these ten requirements, we investigate recent five schemes. Juang et al.'s scheme suffers smart card lost and impersonation attack. Kim et al.'s, Hsiang et al.'s, and Li et al.'s schemes are subjected to smart card lost and offline password guessing attack. Finally, Xu et al.'s scheme has weakness of insider impersonation attack.

## REFERENCES

[1] H. Y. Chien, C. H. Chen, "A Remote Authentication Preserving User Anonymity," *Proceedings of the 19th International Conference on Advanced Information Networking and Applications* (AINA '05), Vol.2, pp. 245-248, March 2005.

[2] I. E. Liao, C. C. Lee, M. S. Hwang, "A password authentication scheme over insecure networks," *Journal of Computer and System Sciences*, Vol. 72, No. 4, pp. 727-740, June 2006.

[3] T. H. Chen, W. B. Lee, ¡A new method for using hash functions to solve remote user authentication¡, Computers *& Electrical Engineering*, Vol. 34, No. 1, pp. 53-62, January 2008.

[4] C. S. Bindu, P. C. S. Reddy, B. Satyanarayana, ¡Improved remote user authentication scheme preserving user anonymity¡, *International Journal of Computer Science and Network Security*, Vol. 8, No. 3, pp. 62-65, March 2008.

[5] Y. Lee, J. Nam, D. Won, ¡Vulnerabilities in a remote agent authentication scheme using smart cards¡, *LNCS: AMSTA*, Vol. 4953, pp. 850-857, April 2008.

[6] W. S. Juang, S. T. Chen, H. T. Liaw, ¡Robust and efficient password-authenticated key agreement using smart cards¡, *IEEE Transactions on Industrial Electronics*, Vol. 55, No. 6, pp. 2551-2556, June2008.

[7] W. S. Juang, W. K. Nien, ¡Efficient password authenticated key agreement using bilinear pairings¡, *Mathematical and Computer Modelling*, Vol. 47, No. 11-12, pp. 1238-1245, June 2008.

[8] J. Y. Liu, A. M. Zhou, M. X. Gao, ¡A new mutual authentication scheme based on nonce and smart cards¡, *Computer Communications*, Vol. 31, No. 10, pp. 2205-2209, June 2008.

[9] T. Xiang, K. Wong, X. Liao, ¡Cryptanalysis of a password authentication scheme over insecure networks¡, *Computer and System Sciences*, Vol. 74, No. 5, pp. 657-661, August 2008.

[10] G. Yang, D. S. Wong, H. Wang, X. Deng, ¡Two-factor mutual authentication based on smart cards and passwords¡, *Journal of Computer and System Sciences*, Vol. 74, No. 7, pp.1160-1172, November 2008.

[11] T. Goriparthi, M. L. Das, A. Saxena, ¡An improved bilinear pairing based remote user authentication scheme¡, *Computer Standards & Interfaces*, Vol. 31, No. 1, pp. 181-185, January 2009.

[12] H. S. Rhee, J. O. Kwon, D. H. Lee, ¡A remote user authentication scheme without using smart cards¡, *Computer Standards & Interfaces*, Vol. 31, No. 1, pp. 6-13, January 2009.

[13] Y. Y. Wang, J. Y. Liu, F. X. Xiao, J. Dan, ¡A more efficient and secure dynamic ID-based remote user authentication scheme¡, *Computer Communications*, Vol. 32, No. 4, pp. 583-585, March 2009.

[14] H. C. Hsiang, W. K. Shih, ¡Weaknesses and improvements of the Yoon¡Ryu¡Yoo remote user authentication scheme using smart cards¡, *Computer Communications*, Vol. 32, No. 4, pp. 649-652, March 2009.

[15] D. Z. Sun, J. P. Huai, J. Z. Sun, J. X. Li, ¡Cryptanalysis of a mutual authentication scheme based on nonce and smart cards¡, *Computer Communications*, Vol. 32, No. 6, pp. 1015-1017, April 2009.

[16] S. K. Kim , M. G. Chung, ¡More secure remote user authentication scheme¡, *Computer Communications*, Vol. 32, No. 6, pp. 1018-1021, April 2009.

[17] H. R. Chung, W. C. Ku, M. J. Tsaur, ¡Weaknesses and improvement of Wang et al.'s remote user password authentication scheme for resource-limited environments¡, *Computer Standards & Interfaces*, Vol. 31, No. 4, pp. 863-868, June 2009.

[18] J. Xu, W. T. Zhu, D. G. Feng, ¡An improved smart card based password authentication scheme with provable security¡, *Computer Standards & Interfaces*, Vol. 31, No. 4, pp. 723-728, June 2009.

[19] M. S. Hwang, S. K. Chong, T. Y. Chen, ¡DoS-resistant ID-based password authentication scheme using smart cards¡, *Journal of Systems and Software*, In Press, Available online 12 August 2009.

[20] C. T. Li, M. S. Hwang, ¡An efficient biometrics-based remote user authentication scheme using smart cards¡, *Journal of Network and Computer Applications*, Vol. 33, No. 1, pp. 1-5, January 2010.

AUTHORS PROFILE

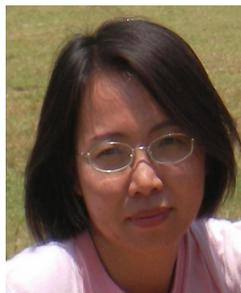

**Yalin Chen** received her bachelor degree in the depart. of computer science and information engineering from Tamkang Univ. in Taipei, Taiwan and her MBA degree in the department of information management from National Sun-Yat-Sen Univ. (NYSU) in Kaohsiung, Taiwan. She is now a Ph.D. candidate of the Institute of Info. Systems and Applications of National Tsing-Hua Univ.(NTHU) in Hsinchu, Taiwan. Her primary research interests are data security and privacy, protocol security, authentication, key agreement, electronic commerce, and wireless communication security.

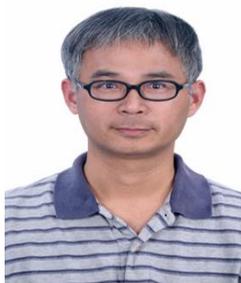

**Jue-Sam Chou** received his Ph.D. degree in the department of computer science and information engineering from National Chiao Tung Univ. (NCTU) in Hsinchu, Taiwan,ROC. He is an associate professor and teaches at the department of Info. Management of Nanhua Univ. in Chiayi, Taiwan. His primary research interests are electronic commerce, data security and privacy, protocol security, authentication, key agreement, cryptographic protocols, E-commerce protocols, and so on.

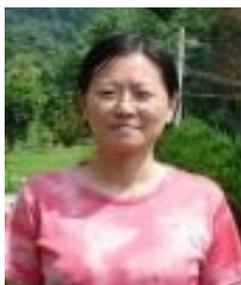

**Chun-Hui Huang** is now a graduate student at the department of Info. Management of Nanhua Univ. in Chiayi, Taiwan. She is also a teacher at Nantou County Shuang Long Elementary School in Nantou, Taiwan. Her primary interests are data security and privacy, protocol security, authentication, key agreement.